\documentclass[12pt,preprint]{aastex}
\usepackage{psfig}
\def\gsim{\lower 2pt \hbox{$\, \buildrel {\scriptstyle >}\over
{\scriptstyle \sim}\,$}}
\def\lsim{\lower 2pt \hbox{$\, \buildrel {\scriptstyle <}\over
{\scriptstyle \sim}\,$}}

\def\xs{{\it CXOGCS J174621.5-285256}}

\shortauthors{Wang}
\shorttitle{X-ray Thread G0.13-0.11}

\begin{document}

\title{X-ray Thread G0.13-0.11: A Pulsar Wind Nebula?}
\author{Q. Daniel Wang, Fangjun Lu, and Cornelia C. Lang}
\affil{Astronomy Department, University of Massachusetts, Amherst, MA 01003,
USA}
\affil{Email: wqd@astro.umass.edu, lufj@flamingo.astro.umass.edu, and
clang@astro.umass.edu}

\begin{abstract}

We have examined {\sl Chandra} observations of the recently discovered 
X-ray thread G0.13-0.11 in the Galactic center Radio Arc region. 
Part of the {\sl Chandra} data was studied by 
Yusef-Zadeh, Law, \& Wardle (2002), who reported the detection of
6.4-keV line emission in this region. We find, however, that this line 
emission is {\sl not} associated with G0.13-0.11. The X-ray spectrum of
G0.13-0.11 is well characterized by a simple power law with an energy slope of 
1.8$^{+0.7}_{-0.4}$ (90\% confidence uncertainties). 
Similarly, the X-ray spectrum of the point-like source embedded in
G0.13-0.11 has a power law energy slope of 0.9$^{+0.9}_{-0.7}$. 
The 2 -- 10 keV band luminosities of these two components 
are $\sim 3.2\times 10^{33}{\rm~ergs~s^{-1}}$ (G0.13-0.11) and 
$\sim 7.5 \times10^{32} {\rm~ergs~s^{-1}}$ 
(point source) at the Galactic center distance of 8 kpc. 
The morphological, spectral, and luminosity properties strongly indicate 
that G0.13-0.11 represents the leading-edge of a pulsar wind nebula, 
produced by a pulsar (point source) moving in a strong magnetic
field environment. The main body of this pulsar wind nebula is likely traced
by a bow-shaped radio feature, which is apparently bordered by G0.13-0.11
and is possibly associated with the prominent nonthermal radio filaments 
of the Radio Arc. We speculate that young pulsars may be 
responsible for various unique nonthermal filamentary radio and X-ray features 
observed in the Galactic center region.

\end{abstract}


\section{Introduction}

The region around the dynamic center of our Galaxy
is very active recently in massive star formation (e.g., Figer et al. 1999),
which should have yielded various high-energy products such as supernova
remnants (SNRs) and neutron stars (e.g., Morris \& Serabyn 1996;
Cordes \& Lazio 1997; Wang, Gotthelf \& Lang 2002). 
Young and fast-rotating neutron stars, in particular, may be 
observable as pulsars, although none is yet known within 
$1^\circ$ radius of the Galactic center (GC). Detection of radio pulsars 
in the GC region is difficult, because of the severe radio wave scattering 
by intervening interstellar ionized gas (Cordes \& Lazio 1997). 
On the other hand, two of the polarized 
radio nonthermal filaments (NTFs; in the Radio Arc and G359.96+0.09) show 
flat or slightly rising positive spectral indices --- a characteristic of 
Crab-like pulsar wind nebulae (PWNe; e.g., Anantharamaiah et al. 1991). 
{\sl Chandra} observations have further revealed various  
diffuse X-ray filaments with unusually hard spectra, which are also signatures
of PWNe. However, no specific link has so far been proposed between such 
radio/X-ray features and PWNe.

Here we report a strong candidate for a PWN that links an X-ray
thread G0.13-0.11 and the prominent NTFs in the Radio Arc region 
(Galactic longitude $l \approx 0^\circ.2$; Fig.\ 1; Yusef-Zadeh, 
Morris, \& Chance 1984). This X-ray thread G0.13-0.11 was
first apparent in the images of Yusef-Zadeh et al. (2002). 
The diffuse X-ray emission from the 
neighboring molecular cloud G0.13-0.13 (Oka et al. 2001) has been
further investigated by Yusef-Zadeh, Law, \& Wardle 
(2002). They considered the X-ray thread as part of a
large-scale diffuse feature that emits the 6.4-keV fluorescent line,
which results from the filling of K-shell vacancies of neutral or
weakly-ionized irons (Koyama et al. 1996; Wang, Gotthelf, \& Lang 2002; 
Wang 2002; Yusef-Zadeh, Law, \& Wardle 2002). Because
 the prominent X-ray thread is on the side of the molecular cloud that 
is opposite to Sgr A$^*$, they concluded that the 6.4-keV line emission could 
not represent the reflection of a possible recent radiation burst from the 
central massive black hole. However, our examination of related {\sl Chandra}
observations shows that the 6.4-keV line is clearly not associated with 
the X-ray thread G0.13-0.11, although the molecular cloud is indeed a 
strong 6.4-keV line emitter (e.g., Fig.\ 2; Wang 2002). We have conducted 
morphological and spectral analyses of both G0.13-0.11 and an embedded 
point-like source. This X-ray study, together with an investigation 
of the radio emission from the region, has led us to conclude that G0.13-0.11 
most likely represents a PWN.

\begin{figure*}[!hbt]
\unitlength1.0cm
    \begin{picture}(16,8) 
\put(-1.1,0){
          \begin{picture}(8,8)
	\psfig{figure=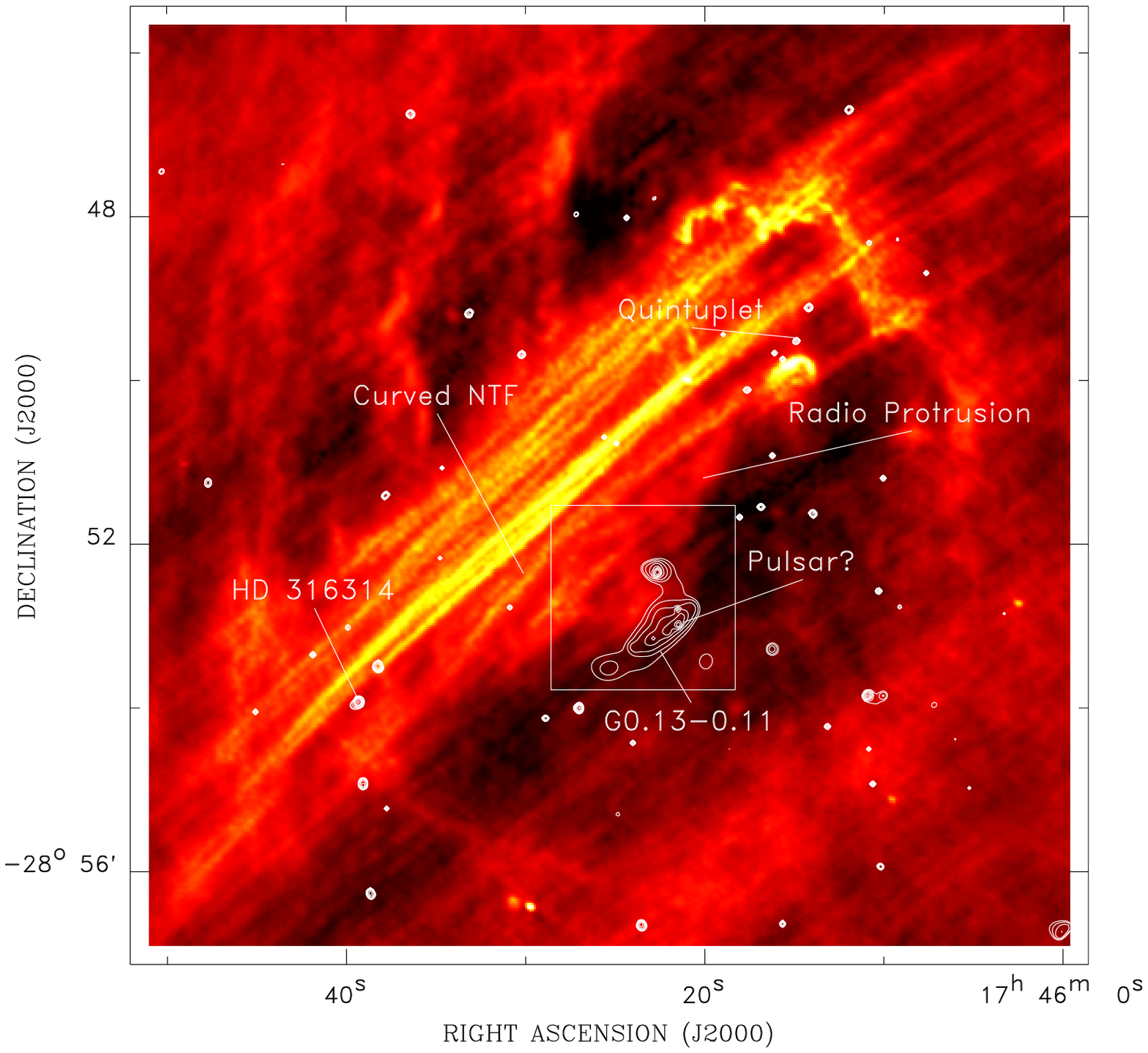,height=8cm,angle=90, clip=}
	\end{picture}
	}
\put(7.8,0){
          \begin{picture}(8,8)
	\psfig{figure=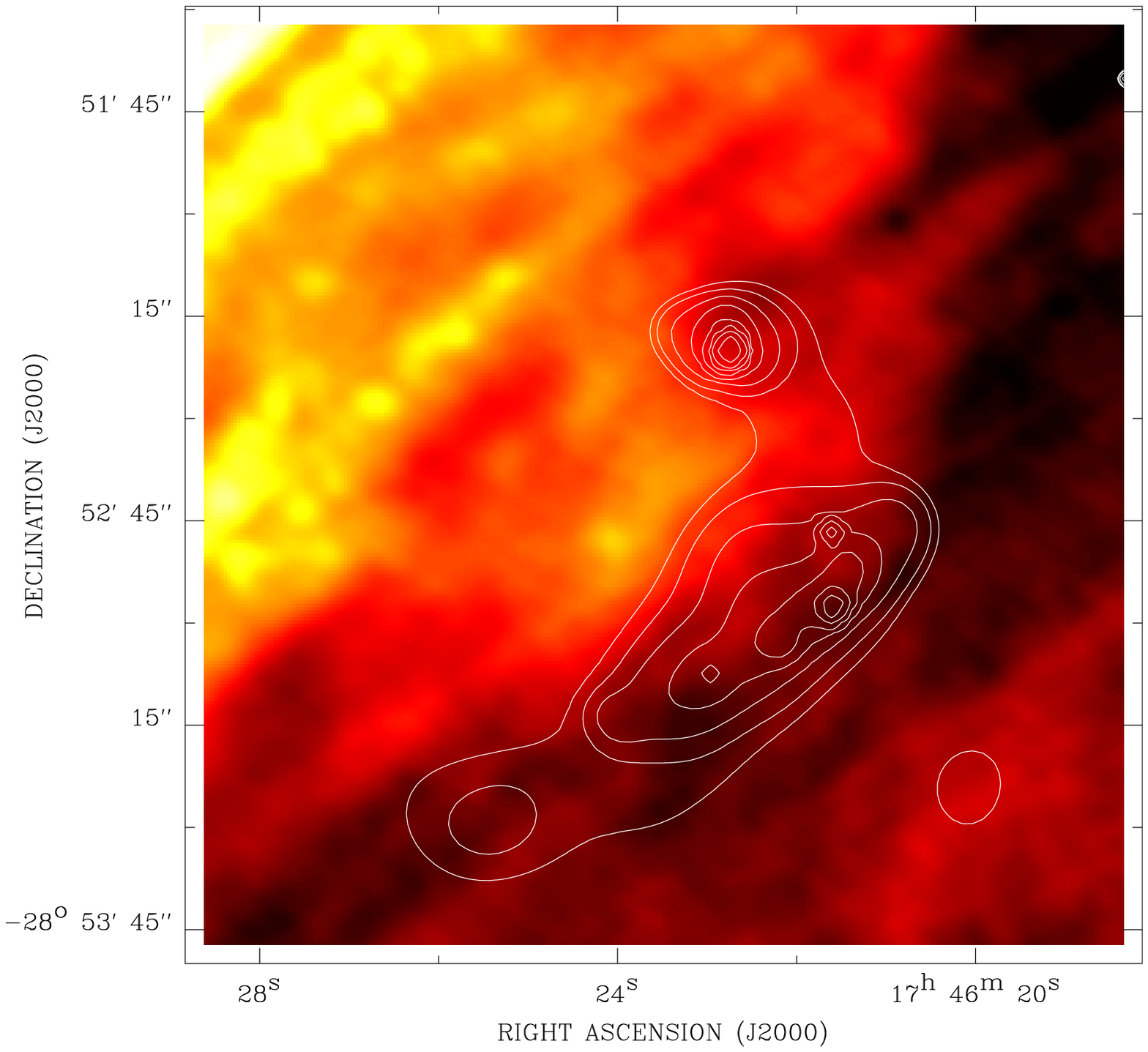,height=8cm,angle=90, clip=}
	\end{picture}
	}
    \end{picture}
\caption{The left panel shows an overview of the region surrounding 
the X-ray thread 
G0.13-0.11: VLA 6-cm radio continuum image (\S 5) and the {\sl Chandra} ACIS-I intensity contours in the 2 -- 6 keV band. The square box illustrates the field of the 
close-up shown in the right panel. The 6-cm image is constructed from the combined CnB and 
DnC data and has a resolution of $\sim$5\arcsec. The X-ray image is adaptively smoothed, using the CIAO routine 
CSMOOTH with the smoothing kernel determined to achieve
a signal-to-noise radio of 2.5-4.
The contours are at 31, 33, 37, 45, 61, 93, 158, and 328 $\times 10^{-3} 
{\rm~counts~s^{-1}~arcmin^{-2}}$. 
}
\end{figure*}

\begin{figure*}[!thb]
\centerline{\psfig{figure=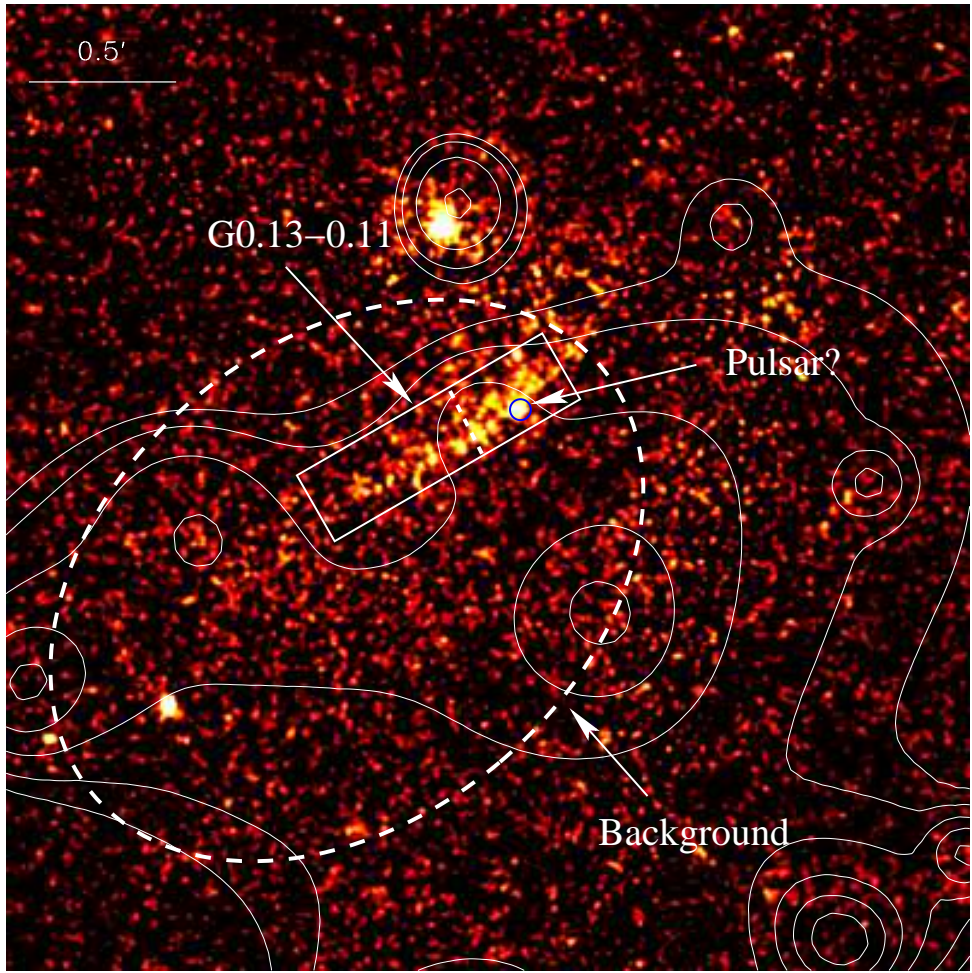,height=6in,angle=0,clip=}}
\caption{A close-up of the region around G0.13-0.11. The image represents 
the 2 -- 6 keV intensity distribution, which has been smoothed with a
Gaussian of size $\sim 1^{\prime\prime}$.
The contours represent the smoothed distribution of 6.4-keV line emission at levels of 0.8, 0.9, 1.1, 1.4, 1.8, and 2.3 $\times
10^{-3} {\rm~counts~s^{-1}~arcmin^{-2}}$ (Wang 2002). Regions for the 
X-ray spectral analysis (\S 3) are also outlined.
}
\end{figure*}

\begin{figure} [!bht]
\centerline{\psfig{figure=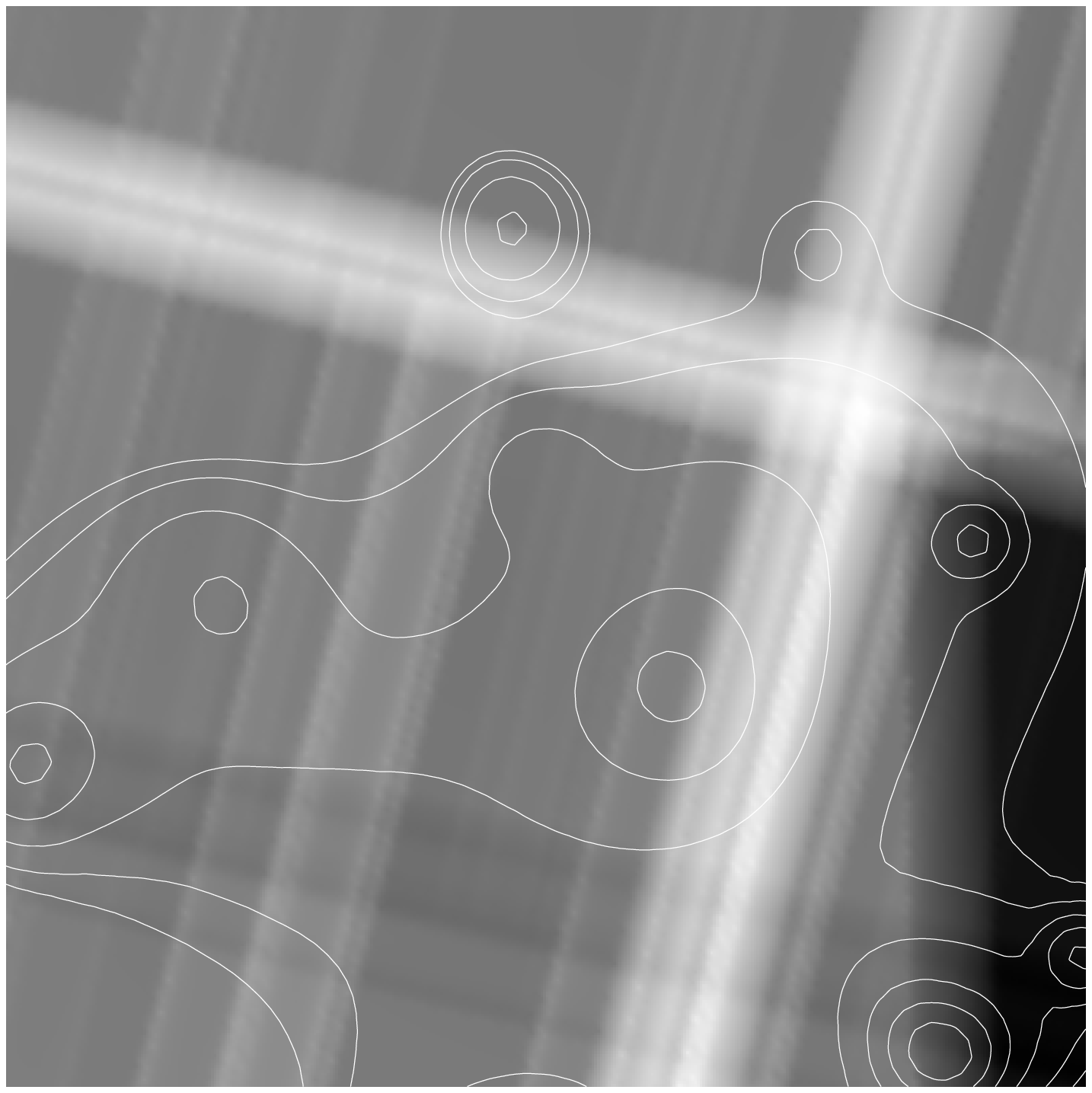,height=6in,angle=90,clip=}}
\caption{ACIS-I effective exposure map in the same field as in Fig.\ 2.
The grey-scale represents the exposure, which ranges from 25 ks to 111 ks.}
\end{figure}

\section{Observations and Astrometry Calibration}

Our study used all existing {\sl Chandra} observations 
of the region around G0.13-0.11; this field is shown in Fig. 1 with
the important features labelled. In addition to 
the cycle-1 observation (50 ks; Obs. ID \#945), which was
used in Yusef-Zadeh et al. (2002) and Yusef-Zadeh, Law, \& Wardle (2002),
we included the recent GC ridge survey data (Obs. ID 
\#2273, 2276, 2282 and 2284 --- $\sim 11$ ks 
each; Wang, Gotthelf, \& Lang 2002) and the 
cycle-1 observation pointed at Sgr A$^*$ (50 ks; Obs. ID \#242). The X-ray data
calibration, including the CTI correction (Townsley et al. 2001), has 
been described by Wang, Gotthelf, \& Lang (2002) and will 
be further detailed by Wang et al. (2002). 

	We searched for X-ray sources in individual observations 
(Wang, Gotthelf, \& Lang 2002). The source position centroids are 
uncertain, both statistically and systematically. The statistical 
uncertainty depends on the count rate of a source and on the point 
spread function, which is a function of the off-axis angle of the 
source in an observation. In order to correct 
for the systematic pointing offset, we select sources with 2$\sigma$ 
statistical uncertainty radii less than $1\farcs6$.

We examined position coincidences between the X-ray sources detected 
in the deep observation (ID \#945) and various optical/near-IR 
objects (e.g., ESO/ST-ECF USNO-A2.0 and 2MASS catalogs).
While no coincidence was found within the 2$\sigma$ statistical 
uncertainty radii of the X-ray source centroids, we considered 
objects projected within the 3$^{\prime\prime}$ radii of the X-ray 
sources. The foreground F0 star, HD316314 
(B=9.94, V=9.51) stands out, which is the brightest optical object within 
the radii. The corresponding point-like X-ray counterpart has 
an ACIS-I count rate of 0.026 ${\rm~counts~s^{-1}}$ and a very soft 
spectrum. A match between the X-ray centroid and the Tycho Reference 
Catalog position of the star ($17^{\rm h}46^{\rm m}39\fs075, 
-28^\circ 53^\prime 51\farcs73$; Hog et al. 1998) requires a shift 
of the X-ray image: 0\farcs27 to the west and 1\farcs00 to the north. 
Making this shift leads to the finding of several other associations between 
optical/near-IR objects and X-ray sources (position coincidences 
within the 2$\sigma$ statistical uncertainty radii). The limiting
R-band magnitude for the selected optical objects, for example, is 17.5.
From the residual
optical/X-ray position offsets of these associations, we infer that 
the positional inaccuracy after the correction has been made should be 
less than $\sim 0\farcs3$. The location of the Quintuplet cluster is 
also indicated in Fig.\ 1. This GC cluster contains numerous massive 
stars including the luminous Pistol Star (Figer et al. 1999). 
However, we detect only one possible X-ray counterpart of a cluster 
member. We find no counterpart for the bright X-ray source just north 
of G0.13-0.11, which is not considered to be part of the X-ray thread 
(right panel in Fig.\ 1).

The focus of this paper is the X-ray thread G0.13-0.11 that borders a 
bow-shaped radio protrusion from the sheath of prominent NTFs and 
surrounds an embedded point-like X-ray source \xs\ 
($17^{\rm h}46^{\rm m}21\fs51, -28^\circ 52^\prime 56\farcs5$ 
with an 1$\sigma$ statistical uncertainty of $\sim 0\farcs15$), marked 
as ``pulsar?'' in Figs.\ 1-2. We have carefully examined possible 
counterparts of this source in other wavelength bands. The nearest 
optical/near-IR object around the X-ray source is U0600-28642574 (B=17.5) at
RA, Dec $= 17^h46^m21^s.81, -28^\circ 52^\prime 56\farcs0$ (J2000), about
4$\farcs$0 from the X-ray centroid. This separation far exceeds the 
position uncertainty of the X-ray source. Furthermore,
the optical and near-IR colors of U0600-28642574
suggest that it is a nearby late-type star (later than M0) with 
little interstellar extinction, which is inconsistent with the large
 column density inferred from the X-ray spectrum (\S 3). We thus conclude 
that \xs\ and U0600-28642574 do not represent the same object.

\section{X-ray Properties of G0.13-0.11}

\begin{figure} 
\centerline{ {\hfil\hfil
\psfig{figure=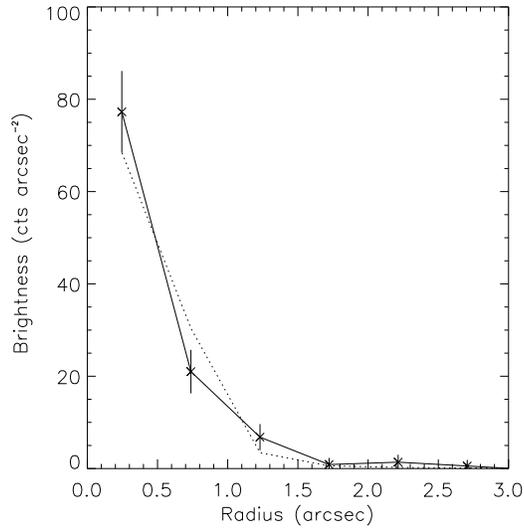,height=3.in,angle=0, clip=}
\hfil\hfil}}
\caption{Comparison of the radial surface brightness 
profile (data points) of \xs\ with the 
{\sl Chandra} ACIS-I point-spread function (dotted line).
}
\end{figure}

Fig.\ 2 presents a close-up of the G0.13-0.11 field. The X-ray thread G0.13-0.13 extends $\gtrsim 
40^{\prime\prime}$ to the southeast from the source \xs\ and has an
intrinsic width of $\sim 2^{\prime\prime}$.
Fig.\ 3 shows potential instrumental effects in the same region as in Fig.\ 2. 
The morphology and intensity 
of the northwestern part of G0.13-0.11, in particular, are weakly constrained because 
of the poor counting statistics within a gap between CCD chips of the 
observation \#945. The intrinsic morphology of G0.13-0.11 could be 
quite symmetric about the source, and thus more extensive to the northwest. Fig.\ 4 compares the radial surface 
brightness profile of the source and the telescope point spread 
function at the position of the observation \# 945. It is clear that the source is 
point-like and cannot simply be a bright clump of G0.13-0.11. 

	Fig.\ 5 shows the spectra of both G0.13-0.11 and \xs.  We extracted 
the spectral data from the observation \# 945, which was also used
 by Yusef-Zadeh, Law \& Wardle (2002). As shown in Fig.\ 2, the extraction 
radius for \xs\ is 1\farcs5, and a rectangular region is used for G0.13-0.11. 
The 6.4-keV line is completely absent in the X-ray spectra of both sources. 

\begin{figure} 
\centerline{ {\hfil\hfil
\psfig{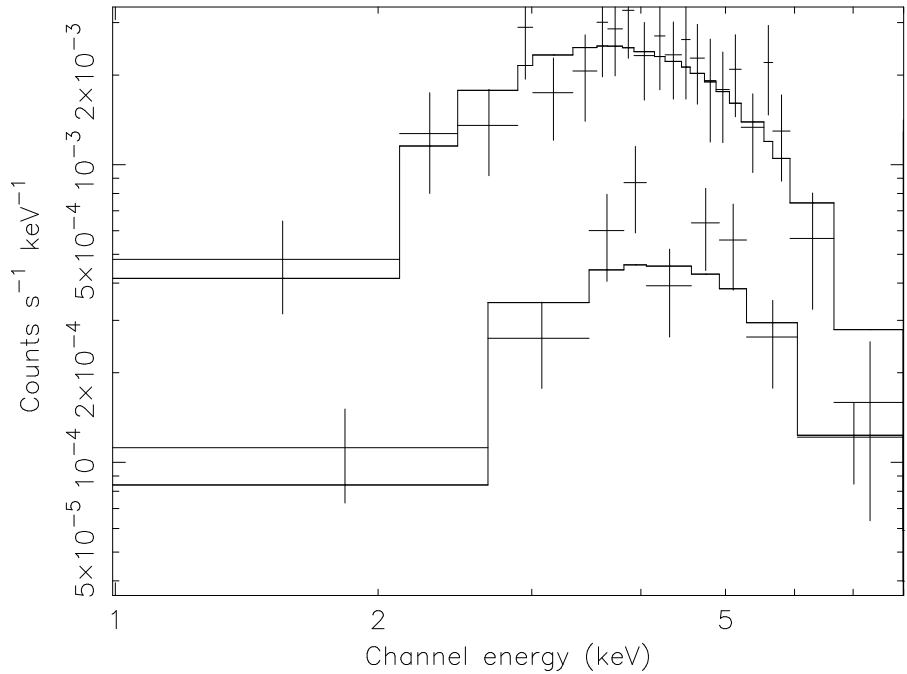}
\hfil\hfil}}
\caption{ACIS-I spectra of the X-ray thread G0.13-0.11 (above) and 
\xs\ (below). 
The histograms represent the best-fit power law models of the data.
}
\end{figure}

While the limited quality of the X-ray data alone does not allow for 
a serious testing of spectral models, we consider various possible
origins of the X-ray emission. For example, although
an one-temperature Raymond-Smith thermal plasma model gives a statistically 
acceptable fit, the inferred temperature has a 99$\%$ lower limit of 8 keV,
which is too high to be consistent with a thermal supernova 
remnant interpretation. Assuming a two-temperature 
thermal plasma model does not solve the problem. 
The temperatures, if allowed to change freely,
converge to the same values as high as 64 keV.
If one temperature is fixed at a value of about a few keV, the other
component, assuming a much higher temperature, then dominates
the overall X-ray spectrum. In short, a thermal model with 
a temperature of a few keV has a too steep spectrum to be consistent
with the data. 

Furthermore, to match the observed flat spectrum with a low energy cutoff 
requires large amounts of absorption, which
in turn demands a high emission measure. For instance,
assuming that the plasma has a single temperature of 2 keV and 
that the depth of the X-ray-emitting region is comparable to the width 
($\sim$2${\arcsec}$) of G0.13-0.11, we estimate the absorbing gas column 
density (90\% confidence interval) as 5.0(3.5-6.5)$\times 10^{22}$ cm$^{-2}$ 
and a thermal pressure of $\sim 2 \times 10^{-7}{\rm~dyn~cm^{-2}}$. This 
pressure appears too great for the thermal gas to be confined. Even 
within the brightest NTFs, the dominant magnetic field pressure is 
$\sim 4 \times 10^{-8}{\rm~dyn~cm^{-2}} B_{\rm mG}^2$, where
$B_{\rm mG}$ is the magnetic field strength in units of milli-Gauss 
(e.g., Yusef-Zadeh \& Morris 1987). 
Therefore, both the high surface brightness and the flat spectral 
characteristics strongly suggest that the X-ray emission 
from G0.13-0.11 is nonthermal. 

We have also tested the bremsstrahlung spectral model proposed by 
Valinia et al. (2000). In this case, K-shell vacancies of iron atoms
are produced by collisions with low energy cosmic-ray electrons. 
The model is approximately a combination of a power law with 
an energy slope of 1.4 and a Gaussian line
centered at 6.4 keV. The predicted flux ratio of the 6.4-keV line 
to the 2 -- 10 keV continuum is about 1/26. The spectrum of G0.13-0.11,
however, shows no sign of the  6.4-keV line. We can rule out 
the line-to-continuum flux ratio with a confidence of $\sim 3\sigma$.

A simple power law gives satisfactory fits to the X-ray spectra of 
both the thread G0.13-0.11 and the point-like source \xs\  (Fig.\ 5). 
The best-fit energy slope is 0.9(0.2 -- 1.8) for the source and 
1.8(1.4 -- 2.5) for the thread. 
The jointly-fitted X-ray-absorbing gas 
column density is $5.9 (4.0 - 9.1) \times 10^{22} {\rm~cm^{-2}}$. The column 
densities, if fitted separately for the two spectra, are
consistent with each other, albeit with 
larger uncertainties. There is also statistically marginal evidence for 
the steepening of the power law with the distance from the source. The 
background-removed count flux
ratio of the 4 -- 8 keV band over the 2 -- 4 keV band is 1.24$\pm$0.24 
in the portion of the thread close to the source (the part of 
the rectangular box northwest to the dashed line in Fig.\ 2) 
and 0.92$\pm$0.19 in the 
portion away from the source (southeastern part). We infer the total 
absorption-corrected luminosity as 
$L_x \sim 4.2 (3.2) \times 10^{33}  {\rm~ergs~s^{-1}}$ for the thread and
$\sim 1.9 (7.5) \times10^{32}  {\rm~ergs~s^{-1}}$ for the source in the
0.1 -- 2.4 keV (2 -- 10 keV) band, respectively.

\section{Interpretation: An X-ray-emitting Pulsar Wind Nebula?}

The above morphological and spectral results place tight constraints on the
nature of the point-like source \xs\ and the X-ray thread G0.13-0.11. 
The only satisfactory interpretation that we have found is that these X-ray 
sources represent an isolated young pulsar and its PWN, which are
likely located near the GC, consistent with the X-ray absorption measured. In this scenario, the X-ray emission from the PWN, results naturally
from the synchrotron cooling of shocked pulsar wind particles (electrons 
and positrons).
Using the empirical relation between $L_x$ and the spin-down 
luminosity ($\dot E$) for X-ray-emitting 
pulsars (Becker \& Tr\"umper 1997) or for
pulsar+PWN systems (Seward \& Wang 1998), we estimate $\dot E$ for the 
putative pulsar  
to be a few times $10^{35} {\rm~ergs~s^{-1}}$, typical of an 
X-ray-emitting pulsar. The power law spectra we obtained for these sources
are also characteristic of such a pulsar and its PWN (e.g., Gotthelf \& Olbert 2002). 

The X-ray morphology of our proposed pulsar/PWN system, however, is quite 
unique. We speculate that the slightly-curved linear morphology is due to the 
strong and organized interstellar magnetic field environment of the GC region 
(Lang et al. 1999 and references therein). Studies of the
interaction between the NTFs and molecular clouds indicate 
a magnetic field strength of $\sim 0.1-1$ mG along 
the NTFs (e.g., Yusef-Zadeh \& Morris 1987; Anantharamaiah et al.
1991; Morris \& Serabyn 1996 and references therein), although the magnetic field 
may be weaker in the general interstellar medium. The orientation 
of the magnetic field (traced by radio polarization) in the vicinity of
G0.13-0.11 follows the long axis of the NTFs in the Radio Arc, i.e., 
nearly perpendicular to the Galactic plane (Yusef-Zadeh \& Morris 1987). 
Fig.\ 1 shows that the X-ray thread has a similar orientation, though 
not aligned with any particular NTF. This similarity indicates that 
the PWN may be confined by the local magnetic field.

X-ray emission traces high-energy particles in the PWN, which are most likely 
accelerated at the reverse shock of the pulsar wind and then move
primarily along magnetic field lines. The expansion of the shocked pulsar
wind (unlikely to be freely expanding) should have a velocity smaller than
the thermal velocity of $c/\sqrt{3}$. 
The X-ray-emitting synchrotron lifetime of the particles is short,
 $\tau \sim (1.3 {\rm~yrs}) \epsilon^{-0.5} B_{\rm mG}^{-1.5}$, where
$\epsilon$ is the X-ray photon energy in units of keV. 
From the characteristic angular scale of 40$^{\prime\prime}$ from \xs\ 
to the southeastern end of G0.13-0.11, we infer 
$B \lesssim 0.3$ mG. Furthermore, the balance between the ram-pressure
of the pulsar wind and the magnetic field pressure gives a characteristic
scale of $\sim 0\farcs04 B^{-1}_{\rm mG}[\dot E/(10^{35} 
{\rm~ergs~s^{-1}})]^{1/2}$ at the distance of 8 kpc. Because this 
scale should be smaller than the observed width ($\sim 2^{\prime\prime}$) 
of the X-ray thread G0.13-0.11, we have $B \gtrsim 0.02$ mG.
The required magnetic field thus seems to be quite reasonable for the
general interstellar medium in the GC region.

\section{Radio Emission Possibly Associated with G0.13-0.11}

Our interpretation of the X-ray thread G0.13-0.11 as the leading-edge of a PWN 
also predicts the presence of an extended radio-emitting region.
The radio synchrotron lifetime of an electron is on the order of $\sim 
(2 \times 10^4 {\rm~yrs}) \nu_{\rm GHz}^{-0.5} B_{\rm mG}^{-1.5}$, about 
four orders of magnitude longer than the X-ray-emitting particle lifetime. 
Therefore, the radio emission in such a system represents the accumulation 
of pulsar wind material and traces the overall morphology (and ``history'') 
of the PWN. 

We have obtained and reprocessed archival radio data
of this region originally observed with the VLA at 20 and 6 cm in the DnC and 
CnB array configurations (Yusef-Zadeh et al. 1984, 
Yusef-Zadeh \& Morris 1987) in order to compare these images to the distribution of X-ray emission and to study the spectral properties of the radio emission. 

\begin{figure} 
\centerline{ {\hfil\hfil
\psfig{figure=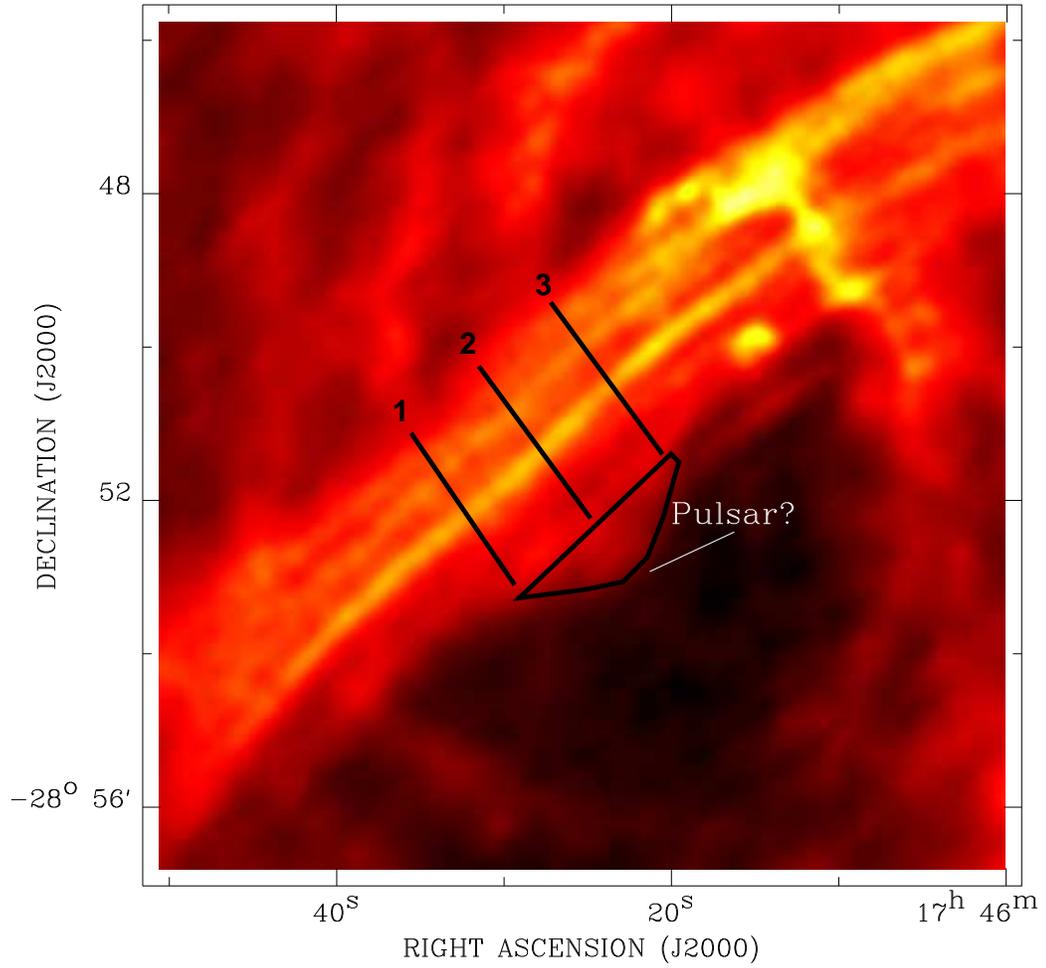,height=6in,angle=270, clip=}
\hfil\hfil}}
\caption{20-cm radio continuum image of the Radio Arc region,
labelled with the three cross-cuts, 
along which the distribution of the 20/6-cm spectral index was 
sampled and is shown in Fig.\ 7. The protrusion of diffuse 
radio emission directly adjacent to G0.13-0.11 is also 
outlined, for which an integrated 20/6-cm spectral 
index of $\alpha$=+0.3 is calculated.}
\end{figure}

\subsection{Radio Morphological Properties} 

An examination of the radio images shows no counterpart of the X-ray source
\xs. This is not a surprise even if 
the source is a young pulsar. The pulsed radio emission 
may be beamed away from us, or the heavy interstellar scattering may have 
significantly diffused the emission (Cordes \& Lazio 1997). 

Fig.\ 1 shows that the curvature of the X-ray thread G0.13-0.11
closely corresponds to the edge of a protrusion of diffuse radio 
emission from the southwestern edge of the prominent Radio Arc NTFs. 
This combination of the features suggests that our proposed PWN is 
produced by a pulsar moving in a strong and organized 
magnetic field environment. Morphologically, G0.13-0.11 differs from 
the Crab Nebula, in which its corresponding pulsar is deeply embedded. 
Instead, G0.13-0.11 shows similarities to ram-pressure confined PWNe, 
such as the Duck/PSR1757-24 (Thorsett et al. 2002) and LMC SNR N157B/PSR 
J0537-6910 (Wang et al. 2001), both of which show offsets of 
X-ray-emitting pulsars from radio peaks. In these latter systems, the X-ray 
emission arises from freshly shocked pulsar winds (thus from regions close to 
the pulsars) whereas radio emission arises from particles swept up 
(into a ``trail'') by the ram-pressure over the lifetime of the nebula. 
However, in the case of G0.13-0.11, the magnetic field likely plays a 
dominant role in confining and transporting shocked pulsar wind particles. 
As a result, a moving pulsar would leave a PWN (a bow-shaped trail of 
shocked wind particles) that expands primarily along the direction of 
magnetic field lines. 

The morphological properties of the protrusion, the NTFs,
and G0.13-0.11 together suggest a likely trajectory of 
the pulsar from the northeast to the present position. 
In this scenario, the synchrotron-emitting relativistic 
particles from the passing pulsar would cause the NTFs in the Radio Arc to 
be illuminated. The NTF (marked as ``curved NTF'' in 
Fig.\ 1) that apparently most adjacent to G0.13-0.11 also shows more 
curvature than any of the other NTFs in the Radio Arc. 
The curvature of the NTFs and G0.13-0.11 may be a result of the balance 
between the magnetic field tension and the thermal pressure of the PWN. 
For the pulsar to cross the Radio Arc NTFs 
and the protrusion ($\sim 4^\prime$ or 10 pc at the GC) requires $\sim 10^5 
{\rm~years} (v_p/100 {\rm~km~s^{-1}})^{-1}$, where $v_p$ is 
the pulsar proper motion speed. Interestingly, this time is consistent with 
the spin-down age ($10^4-10^{6}$ years) of the pulsar expected from the 
observed X-ray luminosity (Becker \& Tr\"umper 1997), and may be compared to
the radio synchrotron lifetime of the particles.

\subsection{Radio Spectral Index}

\begin{figure} 
\centerline{ {\hfil\hfil
\psfig{figure=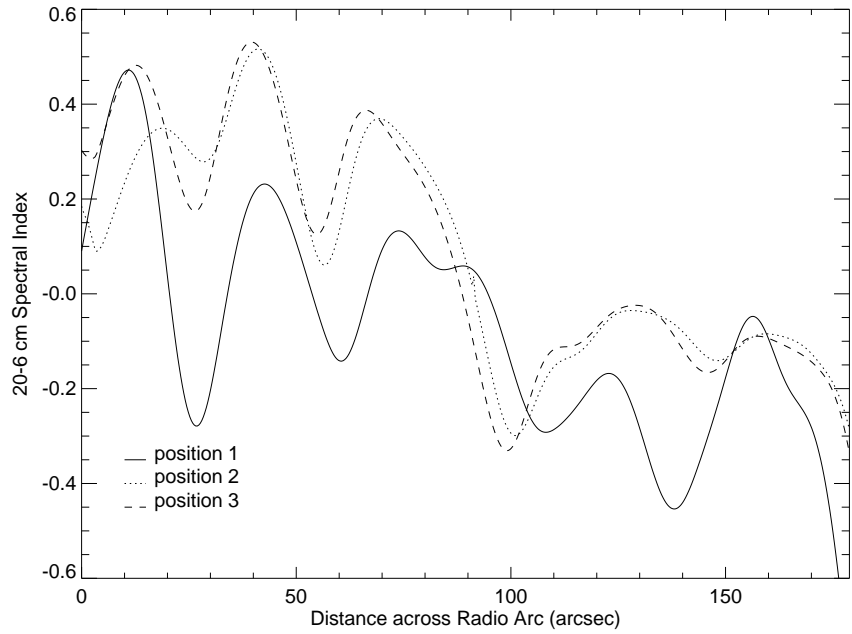,height=3.5in,angle=90, clip=}
\hfil\hfil}}
\caption{Distribution of spectral index for the three cross-cuts 
of the Radio Arc; sampled from the southwest to the northeast. 
The peaks represent the emission from the NTFs of the Radio Arc, 
whereas the valleys represent the weaker, diffuse component of 
radio emission present between the NTFs.}
\end{figure}

We have examined the radio spectral index in the NTFs and 
in the protrusion of radio emission just below the 
NTFs (Fig.\ 6). We have obtained comparable spatial resolutions 
($\sim$ 10\arcsec) by using the 20-cm CnB array data and the 6-cm DnC data. 
One concern, of course, is that the lack of total radio power in these 
interferometric observations may affect the absolute measurement of the
spectral index. The globally averaged 20/6-cm spectral 
index over a circular region of 2\arcmin~radius centered at 
RA, Dec (J2000) $= 17^h 46^m 30^s, -28^\circ 51^{\prime} 30^{\prime\prime}$.
gives a value of $\alpha$$\sim$+0.1 (where S$_{\nu}$ $\propto$ $\nu^{\alpha}$).
This value is consistent with the average spectral index ($\alpha$$\sim$+0.2$\pm$0.1) 
measured by Reich et al. (1988) for the Radio Arc, using single-dish data. 
Using well-matched interferometric data therefore does not 
seem to significantly bias the spectral index measurement.

While the nature of the flat or slightly positive spectral index remains 
unclear, we speculate that the radio-emitting particles in the Radio Arc
NTFs are not accelerated by shocks, which would yield a
steep (negative index) synchrotron spectrum. In our interpretation, 
the particles represent the thermalized pulsar wind. Because the wind is
believed to be ultra-relativistic and to be thermalized at the reverse shock,
no acceleration is required for radio synchrotron particles.
The thermalization alone would result in a Maxwellian particle 
energy distribution. The particles on the lower 
energy part of this distribution could give rise to a positive 
spectral index of the radio synchrotron emission. This interpretation is
similar to the idea that the positive spectral index could be a result 
of the synchrotron emission from particles of a 
mono-energetic distribution (e.g., Reich et al. 1988). In contrast,
X-ray-emitting synchrotron particles, which should be on the higher energy side
of the Maxwellian distribution, require acceleration in the PWN,
most likely at the reverse shock.

We may further expect to detect the aging of the radio-emitting particles 
in the PWN as a function of the distance from the pulsar. In the 
protrusion of diffuse radio emission directly adjacent to G0.13-0.11 (Fig.\ 6), 
the limited signal-to-noise ratio of the radio data allows for only an 
estimate of the integrated spectral index of $\alpha \sim +0.3$. However,
across the Radio Arc, we investigated the distribution of spectral index
(Figs.\ 6-7). Along all
three cross-cuts, the spectral indices of the NTFs have an average
value of $\sim$ +0.4 at the southwestern edge of the Radio Arc, and steepen to
an average value of $\sim$ -0.1 in the northeastern (i.e. toward the numbers
labelled 1-3) region. This spectral index steepening is consistent with our 
interpretation that the particles are being supplied by the pulsar along 
its northeast-to-southwest trajectory. 

\section{Conclusions}

We have studied the X-ray thread G0.13-0.11 and the embedded 
point-like X-ray 
source \xs\ as well as the adjacent nonthermal radio emission in the Radio Arc 
region. The morphological and spectral properties of these features
appear to be consistent with our PWN interpretation. We hope that this study 
will stimulate observational and theoretical studies to 
further explore the connection between pulsars and various nonthermal 
features observed in the unique GC environment. 
The ultimate confirmation of the PWN interpretation requires the detection of the pulsed
signal from the putative pulsar, which could be achieved with 
future radio and/or X-ray observations with fast timing capabilities.

\acknowledgements
We thank the referee B. Aschenbach for valuable comments that helped to 
improve the presentation of the work, which was supported by the SAO grant
G01-2150A.

\end{document}